\documentclass[preprint,
amsmath,amssymb,
aps
]{revtex4-1}
\usepackage{dcolumn}
\usepackage{bm}
\usepackage{caption}
\usepackage{mathrsfs}
\usepackage[titletoc]{appendix}
\usepackage{graphicx}
\usepackage{epstopdf}
\usepackage{hyperref}
\usepackage{subfigure}
\usepackage{slashed}

\begin{document}
	
	\title{Gauge dependence of spontaneous radiation  spectrum of relativistic atomic beam under non-uniform electrostatic field}
	
	\author{Xue-Nan Chen}
	\email{For correspondence: xnchen@hust.edu.cn}
	\affiliation{Hubei Key Laboratory of Gravitational and Quantum Physics, School of Physics, Huazhong University of Science and Technology, Wuhan 430074, China}
	\author{Yu-Hang Luo}
	\affiliation{Hubei Key Laboratory of Gravitational and Quantum Physics, School of Physics, Huazhong University of Science and Technology, Wuhan 430074, China}
	\author{Xiang-Song Chen}
	\email{For correspondence:  cxs@hust.edu.cn}
	\affiliation{Hubei Key Laboratory of Gravitational and Quantum Physics, School of Physics, Huazhong University of Science and Technology, Wuhan 430074, China}
	
	\date{\today}
	\begin{abstract}
		Gauge theory requires physical observables to be gauge-independent. However, ever since Lamb noticed the problem of gauge selection in calculating atomic spontaneous radiation spectrum, the problem of gauge dependence was encountered in many fields of physics research. Therefore, it is important to test the self-consistency of gauge symmetry for various physical systems.  In this paper, we calculate the  transient spontaneous radiation spectrum of  a relativistic hydrogen atom in the non-uniform electrostatic field under the atomic self-reference frame.  The physical system studied in this paper is a frame-transformed version of our recent work [\href{https://link.springer.com/paper/10.1140/epjd/s10053-022-00407-5}{Euro. J. Phys. D \textbf{76}, 84(2022)}] where the radiating object is static while the charge is moving relativistically.  The obtained peak frequency can differ by about $413$ $\mathrm{KHz}$ or larger for the commonly used Coulomb, Lorentz, and multipolar gauges.  This observation can be significant not only for studying how the gauge field interacts with the quantum system in theory, but also for practical experimental applications, such as the timing accuracy of atomic clocks in the external electromagnetic field. 
		
	\end{abstract}

	\maketitle
	\section{Introduction}
	
	The gauge theory constructs the interaction between matter and gauge field, and the self-consistency of gauge theory requires physical observables to be gauge-independent. For classical physics and scattering problems of elementary particles, gauge symmetry can be guaranteed. However, when studying the processes involving quantum bound states, which was first noticed by Lamb in 1952 \cite{Lamb52}, the issue of gauge symmetry seems to be tricky. Lamb found that the experimental results matched well with the results of length gauge $-q\vec E\cdot\vec r$  rather than velocity gauge $-(q/m)\vec A\cdot\vec p$ when studying atomic spectral lines. The problem of gauge dependence also appears in the research fields of  the multi-photon process \cite{Bassani77,Kobe78,Olariu79}, the multi-electron system \cite{Starace71,Anderson74,Lin78,Kobe79}, strong field physics \cite{Bauer06,Faisal07a,Faisal07b,Vanne09}, cavity QED \cite{Dicke54,Rzazewski75,Keeling07,Bernardis18}, and relativistic Coulomb excitation \cite{Bayman05,Dasso06}. Different gauge conditions give different predictions, which contradicts the fact that the experimental results are unique.So people tended to think that there is a kind of ``preferred gauge condition" in these physical research fields \cite{Forney77}. At the same time, many scholars constructed the so-called ``gauge-independent" Hamiltonian operator to solve this conflict \cite{Yang76,Kobe82,Lamb87,Funai19}.
	
	However, the ``gauge-invariant” Hamiltonian operators can currently only applies to the study of quantum systems in weak radiation fields under the dipole approximation. Once the electromagnetic scalar potential is time-varying,the ``gauge-invariant” Hamiltonian operators cannot be applicable. In this case, the Hamiltonian of the system is
	\begin{equation}{\label{HT}}
		H(t)=\frac{1}{2 m_q}(\vec{p}-q \vec{A}(\vec{x}, t))^2+q \phi(\vec{x}, t)+q V(\vec{x}),
	\end{equation}
	Certainly, the ``total”  $H(t)$ is gauge dependent. When using $H(t)$ to define eigenstates and calculate the transition rates, we found they both are gauge dependent. In our recent work \cite{Chen22}, we constructed a quantum system interacting with a relativistic strong Coulomb field to show the inapplicability of  the ``gauge-invariant" Hamiltonian. Although the Schr\"{o}dinger equation under the external field approximation is gauge covariant, there is a problem that the definition of quantum states depends on gauge. Using the external field approximation method, we calculated the transient spontaneous emission spectrum of a one-dimensional quantum harmonic oscillator in the field produced by relativistic moving charge clusters under multipolar gauge, Coulomb gauge, and Lorentz gauge, and found that different gauge conditions give different predictions. We need to use various physical systems to obtain the \textit{effective} external field while an appropriate and full quantum-field method is still lacking \cite{Gross82} and compare the transient radiation spectrum with the result under different gauge.
	
	To test the self-consistency of gauge symmetry, we require that there are relativistic particles in the physical system, and the time-varying scalar potential affects the system adiabatically. Under this experimental demand, the results calculated under different gauge conditions can be well distinguished within the experimental error range. A realistic system is a relativistic atomic beam. In the 1970s, LAMPF could prepare relativistic $ \mathrm {H} ^{-} $ ions, and relativistic atomic beams could be obtained through photodetachment \cite{Bryant71}.   Moreover, with the implementation of ion storage ring projects such as ASTRID \cite{ASTRID}, CRYRING \cite{CRYRING}, ESR \cite{ESR}, TARN II \cite{TARN II}, TSR \cite{TSR}, IUCF \cite{IUCF}, CSR \cite{CSR}, etc., we will get relativistic ion beams more conveniently. Using these experimental facilities, we only need to consider letting relativistic ions or atoms pass through a non-uniform electromagnetic field area to realize our experimental proposal.
	
	In this paper, we adopt the experimental scheme of relativistic atoms  to test the self-consistency of gauge symmetry. The paper is organized as follows: In Sect. \ref{sec:2}, we introduce our physical system: a non-uniform electrostatic field acting adiabatically on a relativistic atomic beam. In the atomic self-reference, parameters can be adjusted to make $\phi_{\text{ext}}(\vec x,t)$ of  the charged ring to be adiabatic for the moving atom. In Sect. \ref{sec:3}, we solve the instantaneous eigen-states of our system by using external field approximation. The solutions indeed differ significantly for the commonly used Coulomb, Lorentz, and multipolar gauges. Then in Sect. \ref{sec:4}, we compute the transient spontaneous radiation spectrum explicitly and find again a significant gauge dependence.  In the last section, we summarize our results and discuss their physical implications.

	\section{A non-uniform electrostatic field acting adiabatically on a relativistic atomic beam}
	\label{sec:2}
	\begin{figure}[htbp]
		\centering
		\captionsetup{justification=raggedright}
		\includegraphics[width=1\linewidth]{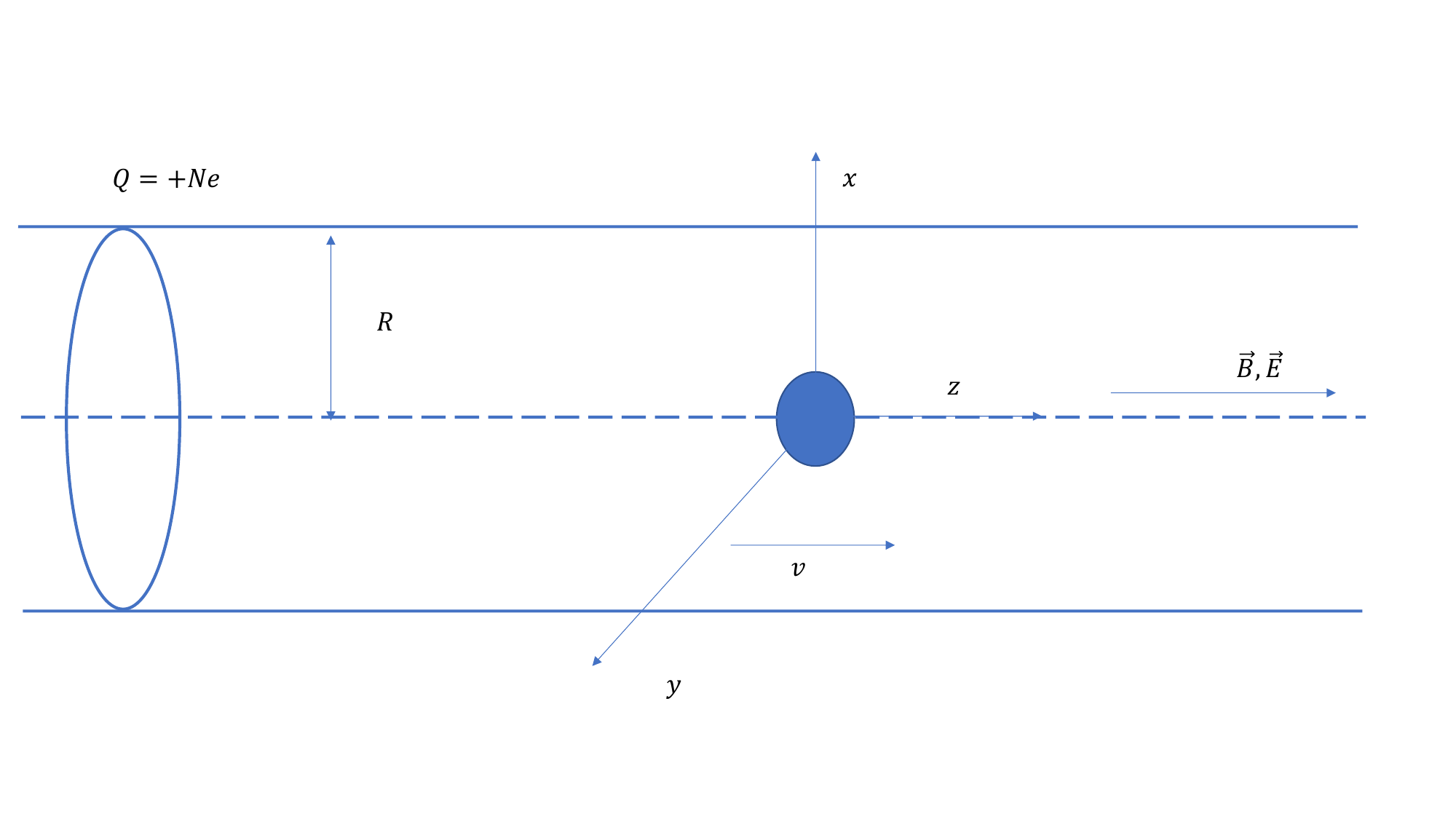}
		\caption{Schematic design of the physical system. The blue solid circle represents a relativistic hydrogen atom with the velocity $v$. The blue coil represents a  charged ring with the charge $+Ne$ and the radius $R$.}
		\label{fig:scheme}
	\end{figure}
	
	Figure \ref{fig:scheme} shows a schematic design of our physical system. The non-uniform electrostatic field is produced by a charged ring with charge $+Ne$ and radius $R$. A relativistic hydrogen atom with velocity $v$ passes through the charged ring. In order to meet the needs of observing transient spontaneous radiation spectrum, we need to meet the adiabatic evolution of excited states. Degenerate quantum states are difficult to meet the requirements of adiabatic conditions, so we need to open the degeneracy of quantum states. To ensure that the energy levels are not degenerate when relativistic atoms pass through the charged ring, we provide a uniform electric and magnetic field along the direction of relativistic atomic motion. In principle, only in the case of non-relativsitic motion, we can use centroid separation method to study the quantum system under the laboratory reference frame \cite{Kaiser17}. In the case of relativistic motion, we can only study the problem in atomic self-reference frame of the quantum system \cite{Bryant83}. So in this paper, we work in the atomic self-reference frame. In the actual data processing, we can transform the experimental results into the atomic self-reference frame. In the atomic self-reference frame, the coordinate system is established with the center of the hydrogen atom as the coordinate origin and the moving direction as the positive direction of the $z$-axis. Because the applied uniform electric field and magnetic fields are along the $z$-axis, the magnitude of the electric and magnetic fields will not change when transforming from the laboratory reference system to the atomic self-reference frame. After the coordinate system is established, the specific expressions of the electric and magnetic fields are $\vec E=(0,0,E)$ and $\vec B=(0,0,B)$ respectively.

	If excited, the relativistic atom can emit a photon by spontaneous radiation. It can be expected that by adjusting  the parameter $N$ and $R$ and selecting the appropriate excited state, in the atomic self-reference frame, the scalar potential $\phi_{\text{ext}}(\vec x,t)$ of the charged ring can be adiabatic for the relativistic atom for duration which is sufficiently long for the excited relativistic atom to emit a transient photon. The velocity of the relativistic atom $v$ should be chosen large enough to produce a considerable difference for $\phi_{\text{ext}}(\vec x,t)$ among various gauges, and at the same time small enough to allow for a rough external-field approximation. Under the external-field approximation method, the Hamiltonian describing the physical system of our schematic design is 
	\begin{equation}{\label{systemH}}
		H_G(t)=\frac{1}{2 m_e}\left[\vec{p}+e \frac{1}{2}(\vec{B} \times \vec{x})+e \vec{A}_{\text{ext}}(\vec{x}, t)\right]^2+V(\vec{x})+e E z-e \phi_{\text {ext }}(\vec{x}, t),
	\end{equation}
	where,  $V(\vec x)=-e^2/r$ is the binding potential energy of hydrogen atom. $(\phi_{\text{ext}}(\vec x,t),\vec A_{\text{ext}}(\vec x,t))$ is the gauge potential generated by a charged ring in the atomic self-reference frame.
	
	\section{The instantaneous eigen-states of the  relativistic atom in atomic self-reference frame}
	\label{sec:3}
	To compute the radiation spectrum of the relativistic atom , one must first define its eigen-states. In our case, this is done by solving
	\begin{equation}{\label{eigns}}
		H_G(t)|\alpha_G(t)\rangle=W_{\alpha G}|\alpha_G(t)\rangle,
	\end{equation}
	According to \cite{Chen22}, We may always apply to Eq. (\ref{eigns}) the following unitary transformation:
	\begin{equation}{\label{trans}}
		|\alpha_G(t)\rangle=U(t)|\alpha^{\prime}_G(t)\rangle,\quad H_G(t)=U(t)H^{\prime}_G(t)U^{-1}(t),
	\end{equation}
	then, Eq. (\ref{eigns}) becomes
	\begin{equation}{\label{eignst}}
		H^{\prime}_G(t)|\alpha^{\prime}_G(t)\rangle=W_{\alpha G}|\alpha^{\prime}_G(t)\rangle,
	\end{equation}
	where the new operator is expressed as
	\begin{equation}{\label{Hp}}
		H^{\prime}_G(t)=\frac{1}{2 m_e}\left[\vec{p}+ \frac{e}{2}\left(\left(\vec{B}+\vec{B}_{\mathrm{ext}}(0, t)\right) \times \vec{x}\right)\right]^2+V(\vec{x})+e E z-e \phi_{\mathrm{ext}}(\vec{x}, t),
	\end{equation}
	when choosing appropriate unitary transformation \(U(t)\). Notice that unitary transformation and operator $H^{\prime}_G(t)$ are merely a mathematical technique, after fixing a gauge, for the convenience of solving the eigen-equation. Hence, when solving Eq. (\ref{eigns}) of the relativistic atom, the vector potential $\vec A_{\text{ext}}(\vec x,t)$ is trivial because after unitary transformation,  vector potential $\vec A_{\text{ext}}$ is transformed into the gauge-invariant form $(1/2)\vec B_{\text{ext}}\times\vec x$. And only the scalar potential $\phi_{\text{ext}}(\vec x,t)$ is essential. 
	
	In our schematic physical system, the relativistic atomic beam moves along the symmetry axis of the charged ring, Hence, \(\vec{B}_{\text {ext }}(0, t)=0\).  When the uniform magnetic field $\vec B$ is weak, we can ignore the effect of the \(\vec{B}^2\) term. For the effect of the external electric field, we only consider the effect of the electric dipole. Hence, we obtain
	\begin{equation}{\label{Hpp}}
		H_G^{\prime}(t) \approx \frac{\vec{p}^2}{2 m_e}+V(\vec{x})+\frac{e}{ 2m_e} B_z l_z+e E z-e(\phi_{\text {ext }}(0, t)+\partial_z \phi_{\text{ext}}(\vec{x}, t)|_{\vec{x}=0} z),
	\end{equation}
	where $l_z$ is the angular momentum operator in $z$ direction of the relativistic atom. In this paper, we mainly study the definition of quantum state under Coulomb, Lorentz, and multipolar gauge. The specific forms and Taylor expansion forms of scalar potential $\phi_{\text{ext}}(\vec x,t)$ of three gauge conditions are shown in Appendix A.
	
	Different from \cite{Chen22} in solving the quantum state of a one-dimensional harmonic oscillator, the solution of the relativistic atom's state affected by non-uniform electrostatic field is difficult to solve by the analytic method. Hence, we use the method of numerical calculation to solve it. We use the eigenstates $|n l m\rangle$ of free Hamiltonian $H_0=\vec p^2/(2m_e)+V(\vec x)$ to expand quantum states
	\begin{equation}{\label{state}}
		\left|\alpha_G^{\prime}(t)\right\rangle=\sum_{n l m} c_{\alpha_G^{\prime}, n l m}(t)|n l m\rangle.
	\end{equation}
	Here, operator $ H _G^{ \prime} (t) $ is expressed as a matrix by using the eigenstate of free Hamiltonian, in which one matrix element is expressed as
	\begin{equation}{\label{matrix}}
		\begin{aligned}
			\left\langle n^{\prime} l^{\prime} m^{\prime}\left|H_G\left(t^{\prime}\right)\right| n l m\right\rangle=&\left(W_n+m \hbar \omega_L-e \phi_{\mathrm{ext}}(0, t)\right) \delta_{n^{\prime} n} \delta_{l^{\prime} l} \delta_{m^{\prime} m}\\&+e\left(E-\left.\partial_z \phi_{\mathrm{ext}}(\vec{x}, t)\right|_{\vec{x}=0}\right)\left\langle n^{\prime} l^{\prime} m^{\prime}|z| n l m\right\rangle,
		\end{aligned}
	\end{equation}
	where, $\omega_L=e B /(2m_e)$. 
	
	By diagonalizing the matrix form of the operator $H_G^{\prime}(t)$, we can obtain the eigenvalues $W_{\alpha G}(t)$ and corresponding eigenvectors $|\alpha_G(t)\rangle$ under the external field approximation. Here, the expansion coefficient $c_{\alpha_G^{\prime}, n l m}(t)$ of the eigenstate is given by each component of the eigenvector obtained by diagonalization method. After some rough estimation, we find that the following values suffice our study: $N=10^{12}$, $\beta=v/c=0.1$, and $R=2.8$ $\mathrm{cm}$.  If the diameter of the charged ring is considered to be $1$ $\mathrm{mm}$, the voltage on the charged ring is in the order of $10^ 5$ $\mathrm{V}$. The intensity of uniform electric and magnetic fields applied in our physical system are $E=1.03\times 10^6$ $\mathrm{V}/\mathrm{m}$ and $B=2.12$ $\mathrm{T}$ respectively. The relativistic atom moves from $z_i=-0.33$ $\mathrm{m}$ to $z_f=0.33$ $\mathrm{m}$. In the following calculation and data expression, we will use the atomic unit system. Hence, $W_n=-1/(2n^2)$ and the operator $H_G^{\prime}(t)$ can be expressed as
	\begin{equation}{\label{mHG}}
		H_G^{\prime}(t)=H_0+E_G(t) H_z+4.5 \times 10^{-6}m,
	\end{equation}
	where, $E_G(t)=E-\left.\partial_z \phi_{\mathrm{ext}}(\vec{x}, t)\right|_{\vec{x}=0}$. $n$, $l$ and $m$ are the principal quantum number, the orbital quantum number, and the magnetic quantum number respectively.  And $H_{z,n^{\prime}l^{\prime}m^{\prime},nlm}=\left\langle n^{\prime} l^{\prime} m'|z| n l m\right\rangle$.
	
\begin{figure}[htbp]
		\centering
		\includegraphics[width=1\linewidth]{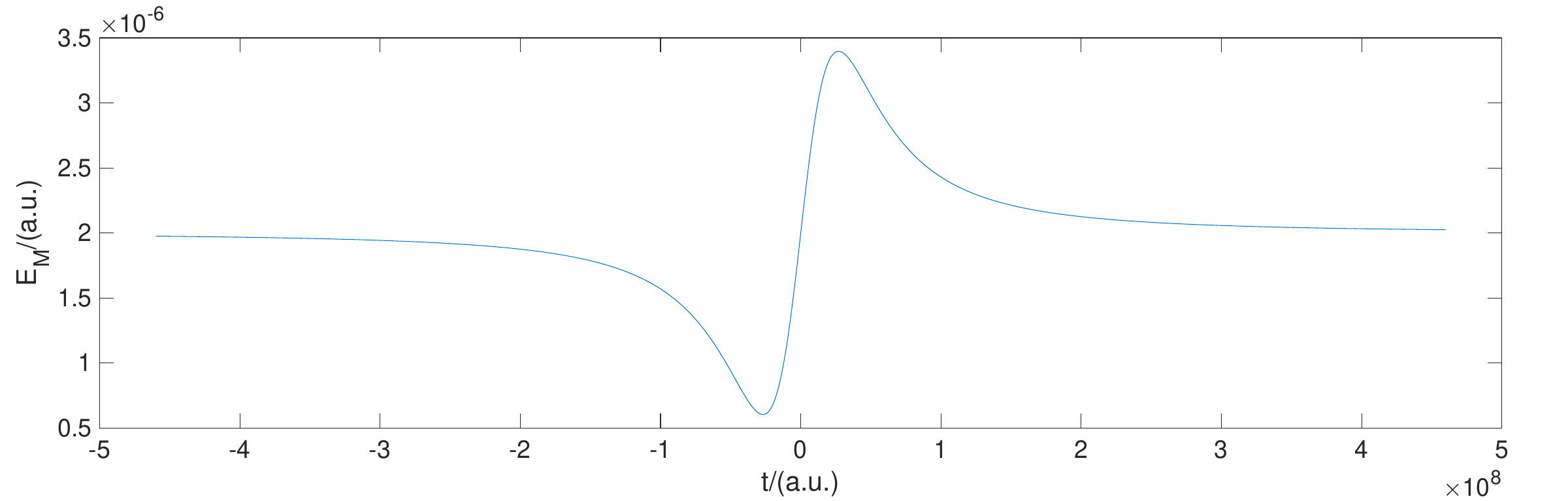}
		\caption{The results of $E_G(t)$ under the multipolar gauge condition.}
		\label{fig:em}
	\end{figure}

\begin{figure}[htbp]
		\centering
		\includegraphics[width=1\linewidth]{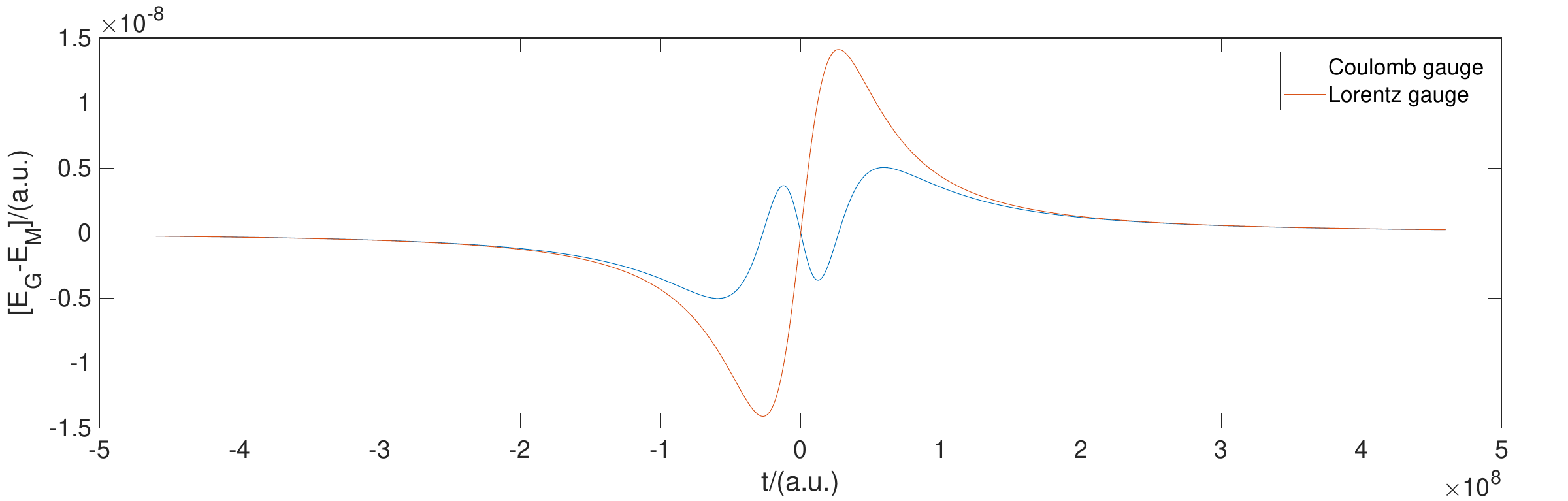}
		\caption{The results of $E_G(t)$ under the Coulomb and Lorentz gauge conditions as differences from the results under the multipolar gauge conditions.}
		\label{fig:eg}
	\end{figure}

	We first show the results of $E_G(t)$ under the multipolar gauge condition with Fig. \ref{fig:em}. Figure \ref{fig:eg} shows the results of $E_G(t)$ under the Coulomb and Lorentz gauge conditions as differences from the results under the multipolar gauge conditions.
	
	The numerical diagonalization method can only deal with finite-dimensional matrices, so we need to truncate the principal quantum number when calculating quantum states and eigenvalues. In this paper, we truncate the principal quantum number at $n=15$.  Because only the electric dipole effect of the non-uniform electrostatic field is considered, the matrix element $E_G(t)H_z$ contributed by the non-uniform electrostatic field is non-zero only when $ l ^{\prime}-l = \pm1 $ and $ m ^{\prime} = m$. Hence, we can use the magnetic quantum number $m$ to divide the matrix $H_z(t)$ and diagonalize it in block. We can obtain the definition of the eigenvector after diagonalization, that is, the quantum state of relativistic atoms in a non-uniform electrostatic field. These quantum states can rank the eigenvalues from small to large under the same magnetic quantum number $m$, and will be given the serial number $N$. In this way, the quantum state can be defined by two parameters, serial number $N$ and magnetic quantum number $m$, namely $|\alpha_G(t)\rangle=|(N,m)_G(t)\rangle$.
	
\begin{figure}[htbp]
		\centering
		\includegraphics[width=1\linewidth]{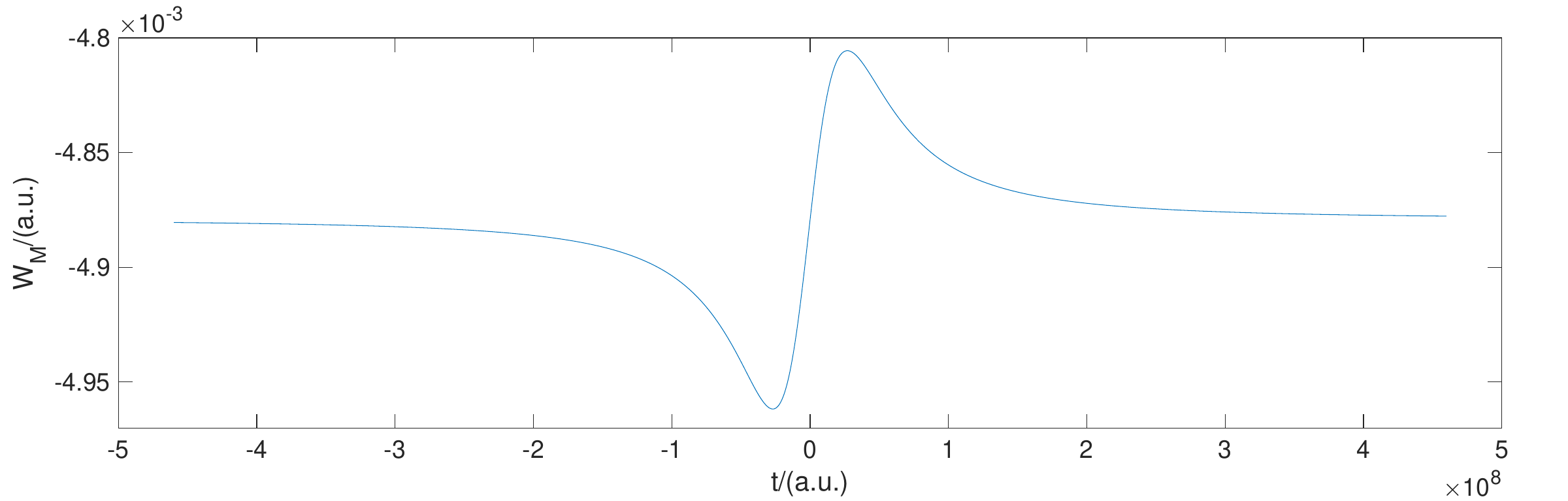}
		\caption{The result of the eigenvalue of  $(53,0)$ under the multipolar gauge condition.}
		\label{fig:enm}
	\end{figure}

\begin{figure}[htbp]
		\centering
		\includegraphics[width=1\linewidth]{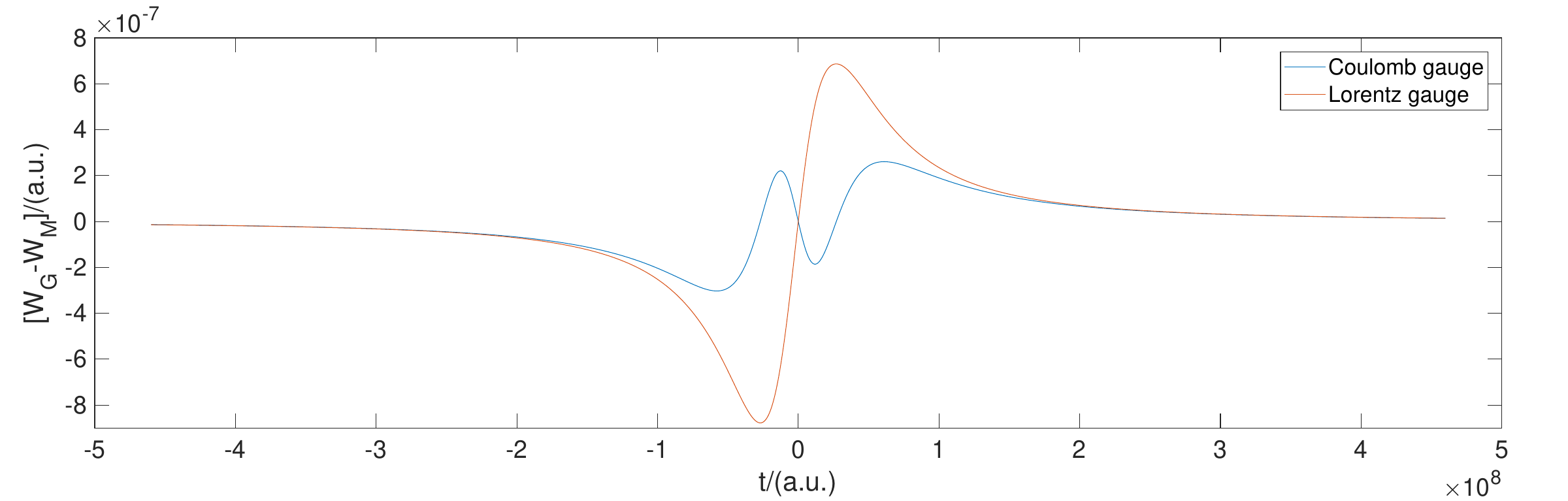}
		\caption{The results of the eigenvalue of $(53,0)$  under the Coulomb and Lorentz gauge conditions as differences from the results under the multipolar gauge condition.}
		\label{fig:eng}
	\end{figure}

	In order to maximize the difference of transient spontaneous emission spectrum under different gauge conditions under the parameters of our physical system, and to satisfy the adiabatic conditions in the following, we choose $(53,0)$ as the excited state to be studied. Figure \ref{fig:enm} shows the result of the eigenvalue of $(53,0)$ under the multipolar gauge condition. Figure \ref{fig:eng} shows the results of  the eigenvalue of $(53,0)$  under the Coulomb and Lorentz gauge conditions as differences from the results under the multipolar gauge condition.
	
	Before solving the eigen-equations, we must check that with our chosen parameters, \(\phi_{\text{ext}}(\vec{x}, t)\) is indeed an adiabatic potential for the relativistic atom, in all three gauges we use. The criterion for the adiabatic approximation is the parameter
	\begin{equation}{\label{adia}}
		r_{\beta \alpha}=\left|\frac{\hbar\langle\beta_G(t) \mid \dot{\alpha}_G(t)\rangle}{W_{\beta G}(t)-W_{\alpha G}(t)}\right| .
	\end{equation}
	where $\alpha,\beta$ label two different states. We evaluate the adiabatic parameters of the quantum state $|\alpha_G(t)\rangle=|(53,0)_G(t)\rangle$. The maximum adiabatic parameters are of the order of $10^{-4}$ under multipole gauge and Coulomb gauge, and of $10^{-3}$ under Lorentz gauge. Thus, if prepared in the state $|(53,0)\rangle$ in the initial moment, the relativistic atom will largely stay in the state $|(53,0)_G(t)\rangle$ during the whole process.
	
	With the adiabatic conditions verified, the gauge dependence of the instantaneous eigenvalues, as we just aforementioned for our designed relativistic atom, may in principle be tested experimentally. A possible experimental observable is the transient spectrum of spontaneous radiation, as we will compute it in the next section.
	
	\section{The transient spontaneous radiation spectrum and its gauge dependence in atomic self-reference frame}
	\label{sec:4}
	We now add into our system the coupling to the background vacuum electromagnetic field. Because in \cite{Chen22}, we find that ``Coulomb-type” gauge dependence is more obvious than ``Lamb-type” gauge dependence, so in this paper, the gauge condition of background electromagnetic field is taken as multipolar gauge. The total Hamiltonian of the system coupled with the vacuum background field is
	\begin{equation}{\label{Ht}}
		H_{\mathrm{tot}}=H_G(t)-q \vec{E}_{\text{vac}} \cdot \vec{r}+H_B
	\end{equation}
	where, $H_G(t)$ is the Hamiltonian of the relativistic atoms under the non-uniform electrostatic field. $E_{\text{vac}}$ is vacuum background field strength
	\begin{equation}{\label{E}}
		\vec{E}_{\text{vac}}=i \sum_{\vec{k}} \sum_\lambda \sqrt{\frac{\hbar \omega_{\vec{k}}}{2 \varepsilon_0 V}}\left(a_{\vec{k}, \lambda} \vec{e}_{\vec{k}} e^{i \vec{k} \cdot \vec{r}}-a_{\vec{k}, \lambda}^{\dagger} \vec{e}_{\vec{k}}^{* \lambda} e^{-i \vec{k} \cdot \vec{r}}\right).
	\end{equation}
	Here,
	\begin{equation}{\label{sum}}
		\sum_{\vec{k}} \rightarrow \frac{V}{(2 \pi)^3} \int d^3 k,
	\end{equation}
	and  physically detectable photons should satisfy $\vec{k} \cdot \vec{e}_{\vec{k}}^\lambda=0$.
	The Hamiltonian of the vacuum background field is
	\begin{equation}{\label{HB}}
		H_B=\sum_{\vec{k}} \sum_\lambda \hbar \omega_{\vec{k}}\left(a_{\vec{k}, \lambda}^{\dagger} a_{\vec{k}, \lambda}+\frac{1}{2}\right).
	\end{equation}
	The second term is often omitted in quantum optical calculations. Since the scale of atoms is much smaller than the wavelength scale of the vacuum background field, only dipole interactions 
	\begin{equation}{\label{Ea}}
		\vec{E}_{\mathrm{vac}} \approx i \sum_{\vec{k}} \sum_\lambda \sqrt{\frac{\hbar \omega_{\vec{k}}}{2 \varepsilon_0 V}}\left(a_{\vec{k}, \lambda} \vec{e}_{\vec{k}}-a_{\vec{k}, \lambda}^{\dagger} \vec{e}_{\vec{k}}^{* \lambda}\right) 
	\end{equation}
	need to be considered. By using the eigenstates of relativistic atoms in the non-uniform electrostatic field solved in the previous section, we can rewrite the total Hamiltonian as
	\begin{equation}{\label{Htot}}
		H_{\text{tot}}=\sum_\alpha E_{G, \alpha}(t) \sigma_{\alpha \alpha}^G(t)+\sum_{\vec{k}} \sum_\lambda \hbar \omega_{\vec{k}} a_{\vec{k}, \lambda}^{\dagger} a_{\vec{k}, \lambda}+\sum_{\vec{k}} \sum_\lambda \sum_{\alpha \beta}\left(g_{\vec{k}, \alpha \beta}^{\lambda, G}(t) \sigma_{\alpha \beta}^G(t) a_{\vec{k}, \lambda}+h . c .\right),
	\end{equation}
	where, $\vec{r}^G_{\alpha \beta}(t) \equiv\langle\alpha_G(t)|\vec{r}| \beta_G(t)\rangle$,  $\sigma^G_{\alpha \beta}(t) \equiv|\alpha_G(t)\rangle\langle\beta_G(t)|$ and $g_{\vec{k}, \alpha \beta}^{\lambda, G}(t)=i e\left[\left(\hbar \omega_{\vec{k}}\right) /\left(2 \varepsilon_0 V\right)\right]^{1 / 2}$\\$\left(e_{\vec{k}}^\lambda \cdot r_{\alpha \beta}^G(t)\right)$. For the interaction Hamiltonian, we take the rotating wave approximation
	\begin{equation}{\label{HI}}
		H_I \approx \sum_{\vec{k}} \sum_\lambda \sum_{\alpha \beta}\left(g_{\vec{k}, \alpha \beta}^{\lambda, G}(t) \sigma_{\alpha \beta}^{G+}(t) a_{\vec{k}, \lambda}+g_{\vec{k}, \beta \alpha}^{* \lambda, G}(t) \sigma_{\alpha \beta}^{G-}(t) a_{\vec{k}, \lambda}^{\dagger}\right),
	\end{equation}
	where
	\begin{equation}{\label{sigmap}}
		\sigma_{\alpha \beta}^{G+}(t)=\left\{\begin{array}{rl}
			\left|\alpha_G(t)\right\rangle\left\langle\beta_G(t)\right|, & \text { when } E_{G, \alpha}(t)>E_{G, \beta}(t) \\
			0, & \text { when } E_{G, \alpha}(t)<E_{G, \beta}(t)
		\end{array}\right.,
	\end{equation}
	and
	\begin{equation}{\label{sigmam}}
		\sigma_{\alpha \beta}^{G-}(t)=\left\{\begin{array}{rl}
			0, & \text { when } E_{G, \alpha}(t)>E_{G, \beta}(t) \\
			\left|\alpha_G(t)\right\rangle\left\langle\beta_G(t)\right|, & \text { when } E_{G, \alpha}(t)<E_{G, \beta}(t)
		\end{array} \right..
	\end{equation}
	
	Considering only the single-photon process, the whole state can be approximated as
	\begin{equation}{\label{wf}}
		|\Psi(t)\rangle \approx \sum_\alpha b_\alpha(t)|\alpha_G(t), 0\rangle+\sum_{\alpha, \vec{k}} \sum_\lambda b_{\alpha_G, \vec{k}}^\lambda(t)\left|\alpha(t), \gamma_{\vec{k}}^\lambda\right\rangle,
	\end{equation}
	where, $|\alpha_G(t), 0\rangle=|\alpha_G(t)\rangle\otimes |0\rangle$ and $|\alpha(t), \gamma_{\vec{k}}^\lambda\rangle=|\alpha(t)\rangle\otimes |\gamma^\lambda_{\vec{k}}\rangle$.  $|n_G(t)\rangle$ is the instantaneous eigen-state of the relativistic atom as we constructed in the previous section, and $|0\rangle$, $|\gamma_{\vec{k}}^\lambda\rangle$ are the Fock states of the photon.  In order to derive the coefficients $b_\alpha$ and $b^\lambda_{\alpha, \vec{k}}$, we introduce the dynamic phase $\theta_\alpha=-\int_{t_i}^{t_f} E_\alpha(s) d s$ and adiabatic phase $\gamma_\alpha=i \int_{t_i}^{t_f}\langle \alpha(s) \mid \dot{\alpha}(s)\rangle$ and write
	\begin{equation}{\label{phase}}
		\begin{aligned}
			b_\alpha(t) & =\exp \left[i\left(\gamma_\alpha(t)+\theta_\alpha(t)\right)\right] c_\alpha(t), \\
			b_{\alpha, \vec{k}}^\lambda(t) & =\exp \left[i\left(\gamma_\alpha(t)+\theta_\alpha(t)\right)-i \omega_{\vec{k}}\left(t-t_i\right)\right] c_{\alpha, \vec{k}}^\lambda(t),
		\end{aligned}
	\end{equation}
	Using the state-evolution equation $i \partial_t|\psi(t)\rangle=H_{\text{tot}}(t)|\psi(t)\rangle$ and  neglecting the transition terms which conserve the photon number because the non-uniform electrostatic field adiabatically affects relativistic atoms in atomic self-reference frame, we obtain differential equations for the new coefficients $c_\alpha$ and $c^\lambda_{\alpha, \vec{k}}$ given by
	\begin{equation}{\label{eom}}
		\begin{aligned}
			\dot{c}_\alpha(t)&=-i \frac{1}{\hbar} \sum_{\vec{k}, \lambda} \overline{\sum}_{\beta} g_{\vec{k}, \alpha \beta}^{\lambda, G}(t) \exp \left(i\left(\gamma_\beta(t)+\theta_\beta(t)-\left(\gamma_\alpha(t)+\theta_\alpha(t)\right)\right)-i \omega_{\vec{k}}\left(t-t_i\right)\right) c_{\beta, \vec{k}}^\lambda(t) ,\\
			\dot{c}_{\beta, \vec{k}}^\lambda(t)&=-i \frac{1}{\hbar} g_{\vec{k}, \beta \alpha}^{* \lambda, G}(t) \exp \left(-i\left(\gamma_\beta(t)+\theta_\beta(t)-\left(\gamma_\alpha(t)+\theta_\alpha(t)\right)\right)+i \omega_{\vec{k}}\left(t-t_i\right)\right) c_\alpha(t),
		\end{aligned}
	\end{equation}
	where, $\overline{\sum}_{\beta}$ stands for the sum of all terms satisfying $E_\beta(t)<E_\alpha(t)$. To handle the decay of the excited state, we follow a method similar to the Weisskopf-Wigner approximation \cite{Weisskopf30}  viz.
	\begin{equation}{\label{ca}}
		\begin{aligned}
			\dot{c}_\alpha(t) & =-\frac{1}{\hbar^2} \sum_{\vec{k}, \lambda} \overline{\sum}_{\beta} g_{\vec{k}, \alpha \beta}^{\lambda, G}(t) \exp \left(i\left(\gamma_\beta(t)+\theta_\beta(t)-\left(\gamma_\alpha(t)+\theta_\alpha(t)\right)\right)-i \omega_{\vec{k}}\left(t-t_i\right)\right) \\
			& \left.\times \int_{t_i}^t g_{\vec{k}, \beta \alpha}^{* \lambda, G}\left(t^{\prime}\right) \exp \left(-i\left(\gamma_\beta\left(t^{\prime}\right)+\theta_\beta\left(t^{\prime}\right)-\left(\gamma_\alpha\left(t^{\prime}\right)+\theta_\alpha\left(t^{\prime}\right)\right)\right)+i \omega_{\vec{k}}\left(t^{\prime}-t_i\right)\right) c_\alpha\left(t^{\prime}\right)\right) .
		\end{aligned}
	\end{equation}
	To proceed with the computation in a clearer form, we introduce $\Delta_{G, \alpha \beta}(t)=[E_{\alpha, G}(t) / \hbar-i\langle\alpha_G(t)| \dot{\alpha}_G(t)\rangle]-[E_{\beta, G}(t) / \hbar-i\langle\beta_G(t) |\dot{\beta}_G(t)\rangle]$.  Since the radiation spectrum typically has a peak frequency, we can effectively perform the frequency integration in the range $(-\infty,+\infty)$, yielding
	\begin{equation}{\label{t}}
		\int_{-\infty}^{\infty} d \omega_k e^{i\left(\omega-\omega_{\vec{k}}\right)\left(t-t^{\prime}\right)}=2 \pi \delta\left(t-t^{\prime}\right).
	\end{equation}
	
	Using Eqs. (\ref{ca}) and (\ref{t}), we thereby obtain
	\begin{equation}{\label{ca1}}
		\dot{c}_\alpha(t)=-\sum_\lambda \overline{\sum}_{\beta} \int \frac{\Delta_{G, \alpha \beta}(t)^3 \mathrm{~d} \cos \theta \mathrm{d} \phi}{16 \hbar \pi^2 c^3 \varepsilon_0} e^2\left|\vec{e}_{\vec{k}}^\lambda \cdot \vec{r}_{\alpha \beta}^G(t)\right|^2 c_\alpha(t).
	\end{equation}
	See Appendix B for the detail of the calculation process of $\left|\vec{e}_{\vec{k}}^\lambda \cdot \vec{r}_{\alpha \beta}^G(t)\right|^2$. Hence
	\begin{equation}{\label{ca2}}
		\dot{c}_\alpha(t)=-\overline{\sum}_{\beta} \frac{e^2 \Delta_{G, \alpha \beta}(t)^3\left|\vec{r}^G_{\alpha \beta}(t)\right|^2}{6 \hbar \pi c^3 \varepsilon_0} c_\alpha(t).
	\end{equation}
	Then, to obtain the solutions for \(c_\alpha(t)\) and \(c^\lambda_{\beta, \vec{k}}(t)\), we combine Eq. (\ref{ca2}) with the second line of Eqs. (\ref{eom}), compute the relevant matrix elements, and perform the numerical integration step by step. Ultimately, the to-be-observed radiation spectrum, accumulated until a time \(t_f\), is computed via
	\begin{equation}{\label{S}}
		\begin{aligned}
			\int d \omega_{\vec{k}} S_\beta\left(\omega_{\vec{k}}, t_f\right)&=\sum_{\lambda}\int \frac{d^3 k}{(2 \pi)^3}\left|c^\lambda_{\beta, \vec{k}}\left(t_f\right)\right|^2\\
			&=\sum_\lambda \int \frac{\omega_{\vec{k}}^2 \mathrm{d}\omega_{\vec k} \mathrm{d} \cos \theta \mathrm{d} \phi}{(2 \pi)^3c^3}\left|c_{\beta, \vec{k}}^\lambda(t_f)\right|^2 ,
		\end{aligned}
	\end{equation}
	where, $S_\beta$ is the spontaneous radiation spectrum of the transition process $(53,0)\rightarrow\beta$.  The selection rules of the transition process can be obtained: $\Delta m=0,\pm1$. Here, the process of $\Delta m=0$ is called ``$\pi$-type" spontaneous radiation. The process of $\Delta m=\pm 1$ is called ``$\sigma$-type" spontaneous radiation. According to Eq. (\ref{S}), the results of the spontaneous radiation spectrum of the maximum branching ratio process in the two types can be obtained
	\begin{equation}{\label{S1}}
		S_{\beta}\left(\omega_{\vec{k}}, t_f\right)=\sum_\lambda \int \frac{\omega_{\vec{k}}^2 \mathrm{~d} \cos \theta \mathrm{d} \phi}{(2 \pi)^3c^3}\left|c_{\beta, \vec{k}}^\lambda(t_f)\right|^2 ,
	\end{equation}
	where, $\beta=(43,0)$ or $(34,\pm 1)$.
	
\begin{figure}[htbp]
		\centering
		\includegraphics[width=1\linewidth]{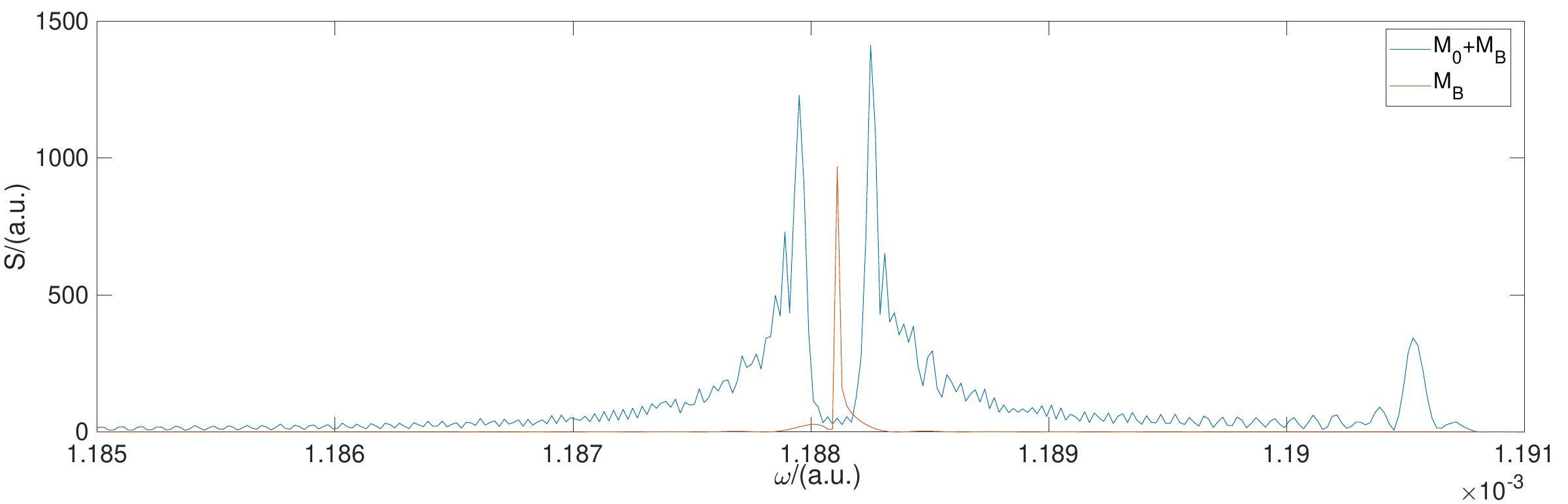}
		\caption{The result of the spontaneous radiation spectrum of the process $(53,0)\rightarrow(43,0)$ under multipolar gauge.}
		\label{fig:sm}
	\end{figure}

\begin{figure}[htbp]
		\centering
		\includegraphics[width=1\linewidth]{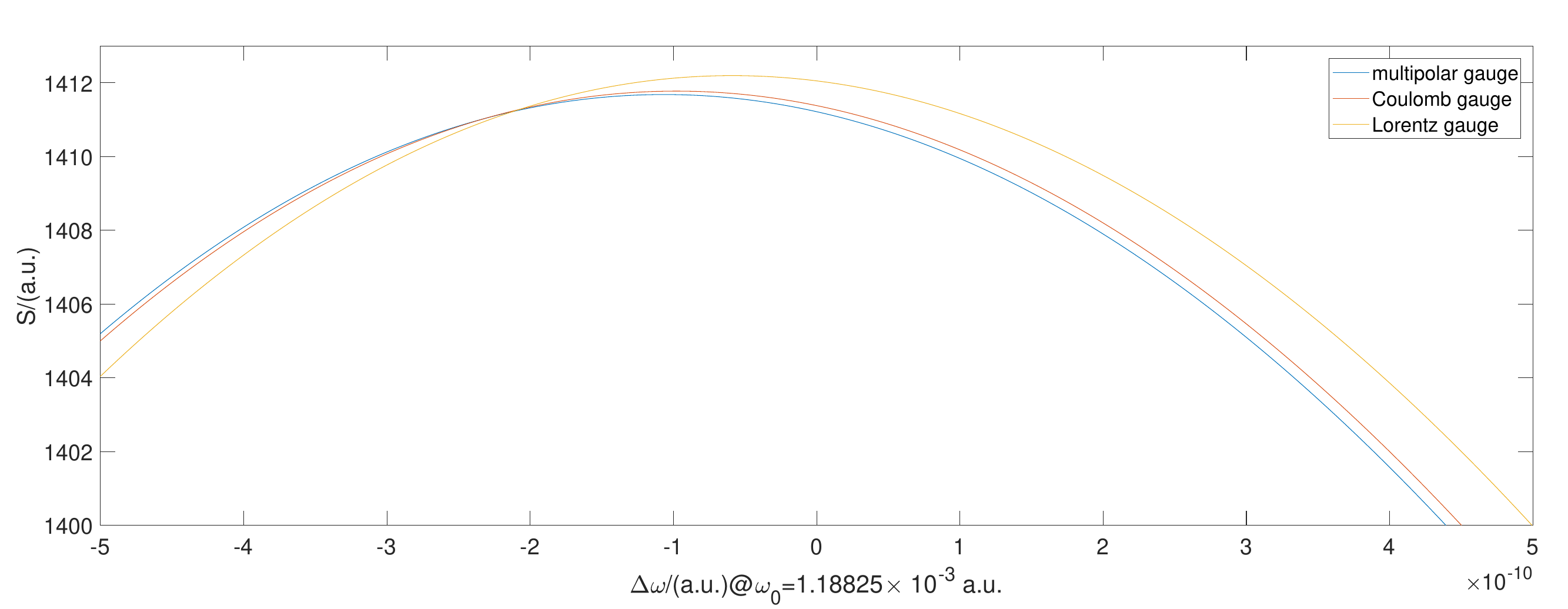}
		\caption{The result of the spontaneous radiation spectrum of the process $(53,0)\rightarrow(43,0)$  under three gauge conditions near the  frequency $\omega_0=1.18825\times 10^{-3}$ $\mathrm{a.u.}$.}
		\label{fig:sg}
	\end{figure}

\begin{figure}[htbp]
		\centering
		\includegraphics[width=1\linewidth]{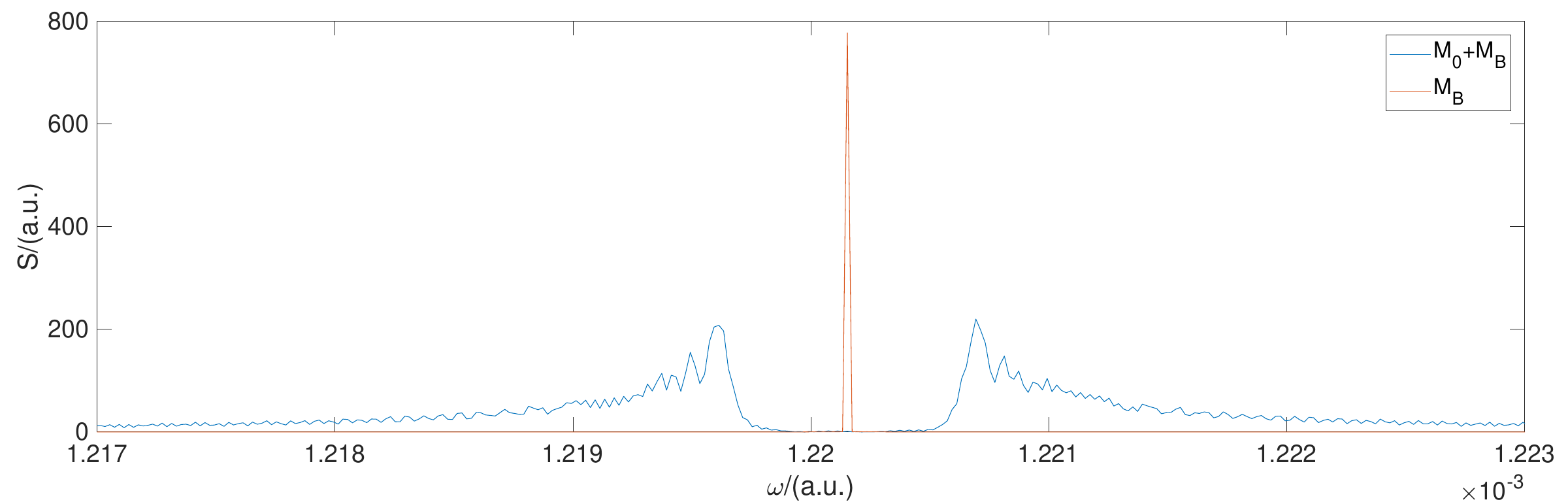}
		\caption{The result of the spontaneous radiation spectrum of the process $(53,0)\rightarrow(34,-1)$ under multipolar gauge.}
		\label{fig:smm}
	\end{figure}

\begin{figure}[htbp]
		\centering
		\includegraphics[width=1\linewidth]{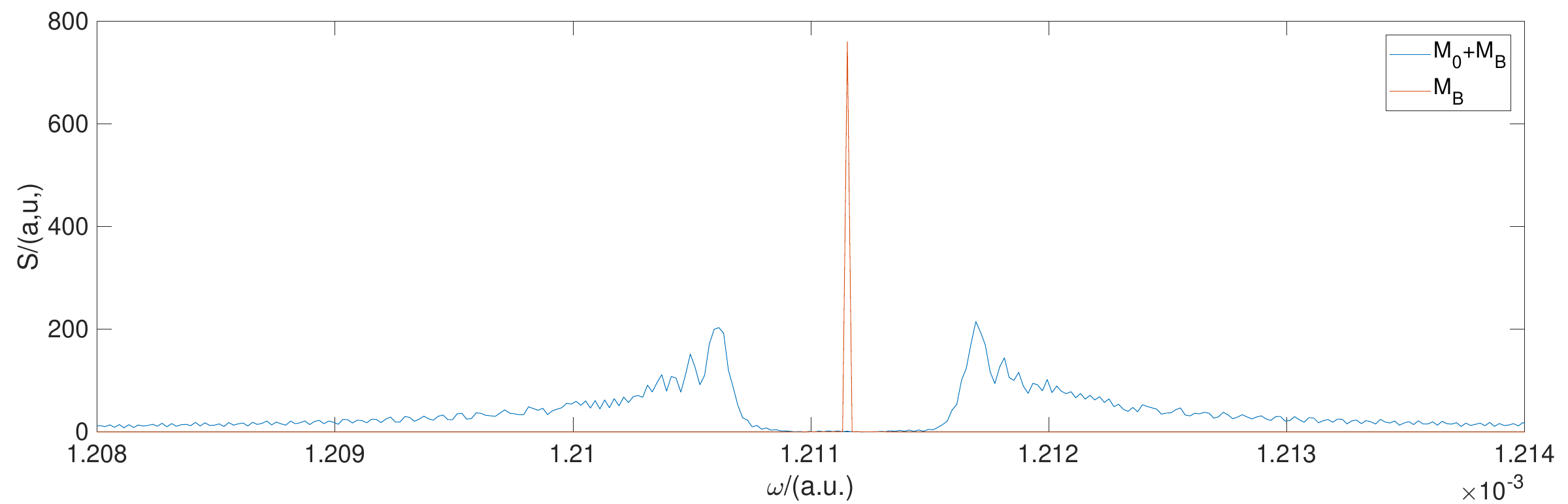}
		\caption{The result of the spontaneous radiation spectrum of the process $(53,0)\rightarrow(34,1)$ under multipolar gauge.}
		\label{fig:spm}
	\end{figure}

\begin{figure}[htbp]
		\centering
		\includegraphics[width=1\linewidth]{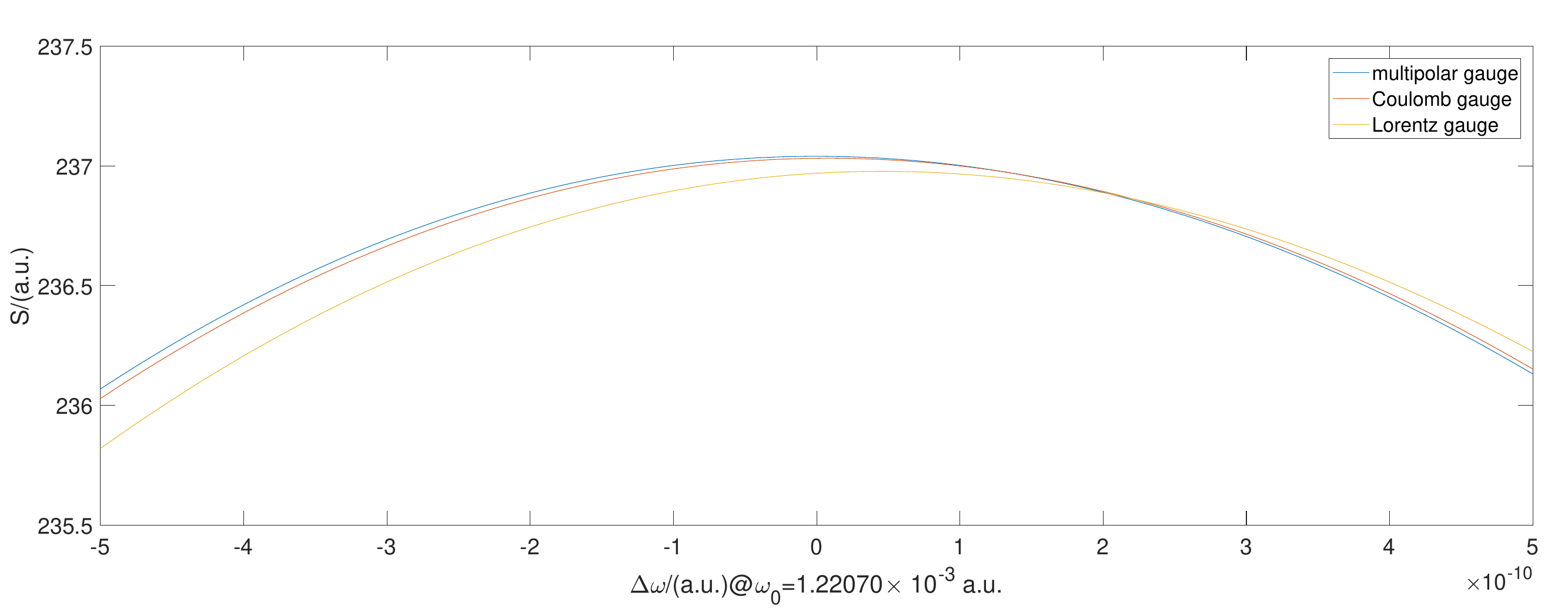}
		\caption{The result of the spontaneous radiation spectrum of the process $(53,0)\rightarrow(34,-1)$  under three gauge conditions near the frequency $\omega_0=1.2207\times 10^{-3}$ $\mathrm{a.u.}$.}
		\label{fig:smg}
	\end{figure}

\begin{figure}[htbp]
		\centering
		\includegraphics[width=1\linewidth]{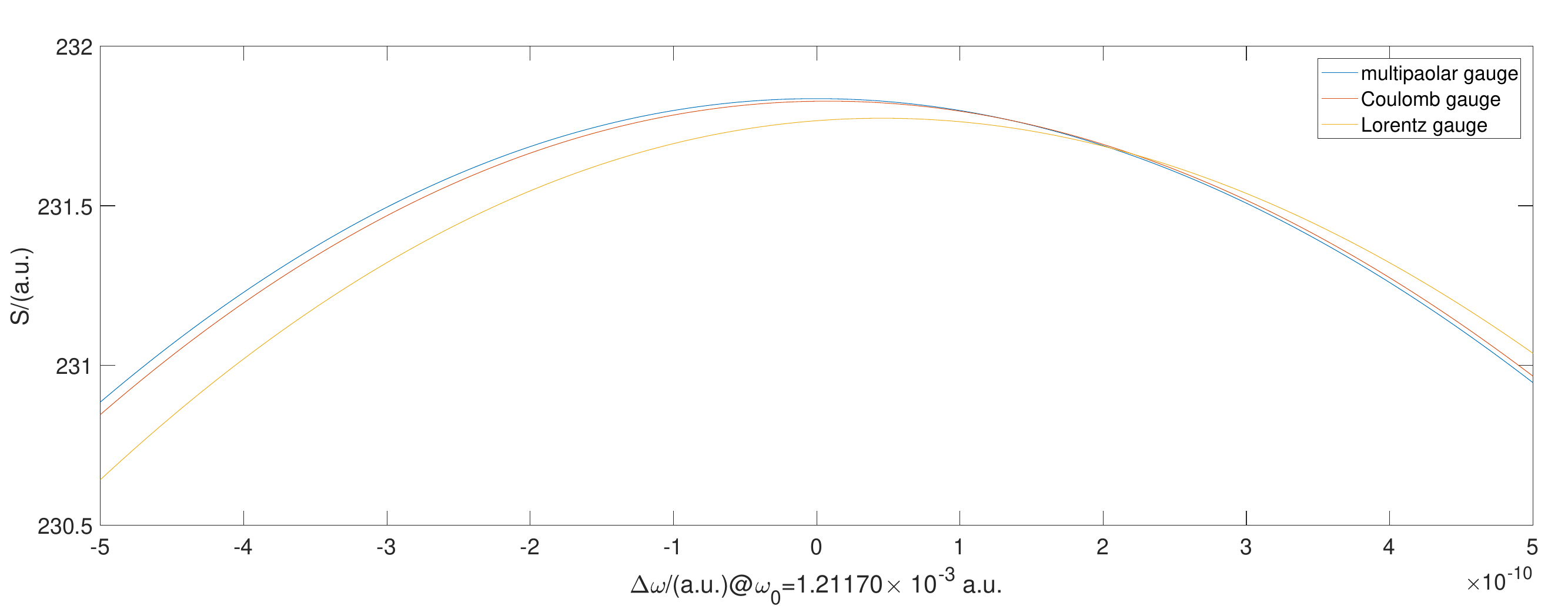}
		\caption{The result of the spontaneous radiation spectrum of the process $(53,0)\rightarrow(34,1)$  under three gauge conditions near the frequency $\omega_0=1.21170\times 10^{-3}$ $\mathrm{a.u.}$.}
		\label{fig:spg}
	\end{figure}

	Figure \ref{fig:sm} shows the result of the ``$\pi$-type" spontaneous radiation spectrum of the maximum branching ratio process $(53,0)\rightarrow(43,0)$ under multipolar gauge. Figure \ref{fig:sg} shows the result of the ``$\pi$-type" spontaneous radiation spectrum of the process $(53,0)\rightarrow(43,0)$ under three gauge conditions near the peak frequency. Figure \ref{fig:smm} and Fig. \ref{fig:spm} show the result of the ``$\sigma$-type" spontaneous radiation spectrum of the maximum branching ratio process $(53,0)\rightarrow(34,\pm 1)$ under multipolar gauge. Figure \ref{fig:smg} and Fig. \ref{fig:spg} show the result of the ``$\sigma$-type" spontaneous radiation spectrum of the process $(53,0)\rightarrow(34,\pm 1)$ under three gauge conditions near the peak frequency. 
	
	
	
	In Fig. \ref{fig:sm}, Fig. \ref{fig:smm}, and Fig. \ref{fig:spm}, we plot the unperturbed spectrum in the absence of non-uniform electrostatic field generated by charged ring. We find that the spontaneous radiation spectrum is obviously different from the spontaneous radiation spectrum of the stationary state. The peak frequency corresponding to the maximum radiation intensity under multipolar gauge is about $1.4 \times 10^{-7}$ $\mathrm{a.u.}$ for ``$\pi$-type" and about $5.4 \times 10^{-7}$ $\mathrm{a.u.}$ for ``$\sigma$-type" higher than that of unperturbed spectrum. We also find that under the influence of non-uniform electrostatic field, the radiation intensity of ``$\pi$-type"  is enhanced, while that of ``$\sigma$-type" is significantly weakened.
	
	We compare the peak frequency corresponding to the maximum radiation intensity under three gauge conditions, and find that the peak frequency corresponding to the maximum radiation under the Coulomb gauge is about $9 \times 10^{-12}$ $\mathrm{a.u.}$ for ``$\pi$-type" and about $8 \times 10^{-12}$ $\mathrm{a.u.}$ for ``$\sigma$-type" higher than that under the multipolar gauge and about $3.9 \times 10^{-11}$ $\mathrm{a.u.}$ for ``$\pi$-type" and about $3.8 \times 10^{-11}$ $\mathrm{a.u.}$ for ``$\sigma$-type" lower than that under Lorentz gauge.
	
	\section{Summary and discussion}
	
	In this paper, we proposed a physical system in which a relativistic atomic beam passes through the non-uniform electrostatic field produced by a charged ring. We calculated the spontaneous radiation spectrum of our physical system and found gauge dependence which cannot be cured by existing methods. The obtained peak frequency can differ by about $413$ $\mathrm{KHz}$ or larger for the commonly used Coulomb, Lorentz, and multipolar gauges by using SI units.  Compared with the physical system we studied before \cite{Chen22}, there are mature experimental methods to prepare the physical system we need. This provides a feasible experimental scheme for us to test the self-consistency of gauge theory under the external field approximation method. 
	
	Since we calculate the spontaneous emission in the atomic self-reference frame, one has to transform measurement results from the laboratory reference frame to this frame when comparing the measurement with our calculation. Although there are some risks in defining quantum states by using semi-classical Hamiltonian (\ref{systemH}) obtained by the external field approximation method in the atomic self-reference frame, We can still get the definition of \textit{effective} external field potential by fitting experimental data in the absence of complete quantum field theory. This effective external potential will help us to understand the physical significance of the gauge potential.
	
	Meanwhile, the design we proposed also makes a tunable system to study the relativistic bound-state problems. Because the quantum state of the relativistic atom is defined in the atomic  self-reference frame, in our design, the acting charged ring is relativistic. Moreover, the design makes both the relativity parameter $\beta=v/c$ and the interaction strength continuously adjustable.
	
	Our design also has value in practical application. For example, atomic clocks in satellites are often influenced by the flux of charged particles from the universe. In this case, it has become a realistic problem to determine which gauge should be applied to ensure the accuracy of timing.
	
	\begin{acknowledgments}
		The work had been, and was partly supported by the China NSF via Grants No. 11275077 and No. 11535005. 
	\end{acknowledgments}	
	
	\section*{Author Contribution Statement}
	
	X.-N. Chen and Y.-H. Luo are the co-first authors of this paper. X.-N. Chen and X.-S. Chen are co-corresponding authors of this paper.
	
	\appendix
	
	\section{The scalar potential in different gauges}
	
	For Lorentz gauge,
	\begin{equation}
		\phi_L=\frac{Q}{4\pi\varepsilon_0\sqrt{(z+vt)^2+(1-\beta^2)R^2}}.
	\end{equation}
	For Coulomb guage,
	\begin{equation}
		\phi_C=\frac{Q}{4\pi\varepsilon_0\sqrt{(z+vt)^2+R^2}}.
	\end{equation}
	At $t$, the expanded forms of these two scalar potentials are
	\begin{equation}
		\phi_L\approx\frac{Q}{4\pi\varepsilon_0}[\frac 1{\sqrt{(vt)^2+(1-\beta^2)R^2}}-\frac{(v t)z}{\sqrt{(vt)^2+(1-\beta^2)R^2}^3}+\cdots],
	\end{equation}
	and 
	\begin{equation}
		\phi_C\approx\frac{Q}{4\pi\varepsilon_0}[\frac 1{\sqrt{(vt)^2+R^2}}-\frac{(vt)z}{\sqrt{(\beta t)^2+R^2}^3}+\cdots].
	\end{equation}
	By using PZW transformation \cite{Power59,Wolley71}, we obtain the result of the multipolar gauge
	\begin{equation}
		\phi_M\approx\frac{Q}{4\pi\varepsilon_0}[-\frac{(1-\beta^2)(vt)z}{\sqrt{(v t)^2+(1-\beta^2)R^2}^3}+\cdots].
	\end{equation}

	\section{Calculation of  $\left|\vec{e}_{\vec{k}}^\lambda \cdot \vec{r}_{\alpha \beta}^G(t)\right|^2$}
	
	The constraint relationship between the momentum and the polarization vector of the physical photon state is $\vec k\cdot\vec e^\lambda_{\vec k}=0$. For example, for a photon with momentum $(0,0,k)$, its polarization vector is $\vec e^{\pm}_{(0,0,k)}=\frac 1{\sqrt{2}}(1 ,\pm i,0)$.
	For a photon with arbitrary momentum $k(\sin\theta\cos\phi,\sin\theta\sin\phi,\cos\theta)$, its polarization vector can be determined by the three-dimensional represatation of $\mathrm{SO}(3)$  $R$ is obtained. The matrix representation in $R$ is
	\begin{equation}
		R=\begin{pmatrix}
			\sin\phi	&\cos\theta\cos\phi  &\sin\theta\cos\phi  \\
			-\cos\phi&\cos\theta\sin\phi  & \sin\theta\sin\phi \\
			0&-\sin\theta  & \cos\theta
		\end{pmatrix}.
	\end{equation}
	Hence,
	\begin{equation}
		\vec e^{\pm}_{\vec k}=\vec e^{\pm}_{(0,0,k)}R^T=\frac 1{\sqrt{2}}(\sin\phi\pm i\cos\theta\cos\phi,-\cos\phi\pm i\cos\theta\sin\phi,\mp i\sin\theta).
	\end{equation}
	Then, we obtain
	\begin{equation}
		\vec e^{\pm}_{\vec k}\cdot\vec r^G_{\alpha\beta}(t)=\frac 1{\sqrt{2}}[(\sin\phi\pm i\cos\theta\cos\phi)x^G_{\alpha\beta}(t)+(-\cos\phi\pm i\cos\theta\sin\phi)y^G_{\alpha\beta}(t)\mp i\sin\theta z^G_{\alpha\beta}(t)].
	\end{equation}
	Hence,
	\begin{equation}
		\begin{aligned}
			|\vec e^{\pm}_{\vec k}\cdot\vec r^G_{\alpha\beta}(t)|^2&=\frac 1{\sqrt{2}}[(\sin\phi\pm i\cos\theta\cos\phi)x^G_{\alpha\beta}(t)+(-\cos\phi\pm i\cos\theta\sin\phi)y^G_{\alpha\beta}(t)\mp i\sin\theta z^G_{\alpha\beta}(t)]\\
			&\cdot\frac 1{\sqrt{2}}[(\sin\phi\mp i\cos\theta\cos\phi)x^G_{\alpha\beta}(t)+(-\cos\phi\mp i\cos\theta\sin\phi)y^G_{\alpha\beta}(t)\pm i\sin\theta z^G_{\alpha\beta}(t)]\\
			&=\frac 12[(\sin^2\phi+\cos^2\theta\cos^2\phi)(x^G_{\alpha\beta}(t))^2-2\sin^2\theta\cos\phi\sin\phi(x^G_{\alpha(t)})(y^G_{\alpha\beta}(t))\\
			&+(\cos^2\phi+\cos^2\theta\sin^2\phi)(y^G_{\alpha\beta}(t))^2-2\cos\theta\sin\theta\sin\phi(y^G_{\alpha\beta}(t))(z^G_{\alpha\beta}(t))\\
			&+\sin^2\theta(z^G_{\alpha\beta}(t))^2-2\cos\theta\sin\theta\cos\phi(z^G_{\alpha\beta}(t))(x^G_{\alpha\beta}(t))].
		\end{aligned}
	\end{equation}
	The integral of the cross term coefficient $f(\theta,\phi)$ to the solid angle is $0$, while the integarl of the square term coefficient $f(\theta,\phi)$ to the solid angle is $4\pi/3$.

\end{document}